\documentclass[reprint, amsmath,amssymb, aps, superscriptaddress]{revtex4-1} 
\usepackage{graphicx}% Include figure files
\usepackage{dcolumn}% Align table columns on decimal point
\usepackage{bm}% bold math
\usepackage{color}
\usepackage[dvipdfmx, hypertex]{hyperref}

\begin{document}
\preprint{IPMU16-0008}

\title{Bootstrapping critical Ising model on three-dimensional real projective space }


\author{Yu Nakayama}
\affiliation{Kavli Institute for the Physics and Mathematics of the Universe (WPI),  \\ University of Tokyo, 5-1-5 Kashiwanoha, Kashiwa, Chiba 277-8583, Japan} 

\date{\today}% It is always \today, today,
             %  but any date may be explicitly\input{../../../../Volumes/HD-PSU2/file_reserved/iwasaki_scgt}

\begin{abstract} 
Given a conformal data on a flat Euclidean space, we use crosscap conformal bootstrap equations to numerically solve the Lee-Yang model as well as the critical Ising model on a three-dimensional real projective space. We check the rapid convergence of our bootstrap program in two-dimensions from the exact solutions available.  Based on the comparison, we estimate that our systematic error on the numerically solved one-point functions of the critical Ising model on a three-dimensional real projective space is less than one percent. Our method opens up a novel way to solve conformal field theories on non-trivial geometries.

\end{abstract}
\pacs{Valid PACS appear here}% PACS, the Physics and Astronomy
\maketitle 
% \section{Introduction}\label{sec:1}
The numerical conformal bootstrap in higher dimensional conformal field theories (CFTs) has provided us with the most efficient way to derive critical exponents in yet-to-be analytically solved theories such as the critical Ising model \cite{ElShowk:2012ht}\cite{El-Showk:2014dwa}\cite{Kos:2014bka}\cite{Simmons-Duffin:2015qma} in three dimensions. It also judges (non)-existence of controversial CFTs in various areas of theoretical physics (e.g. \cite{Nakayama:2014lva}\cite{Nakayama:2014yia}\cite{Nakayama:2014sba}).

Almost all of these numerical studies have been aimed at deriving a conformal data, i.e. operator spectrum and operator product expansion (OPE) coefficients, on  a flat Euclidean space (see, however, \cite{Liendo:2012hy}\cite{Gaiotto:2013nva}\cite{Gliozzi:2015qsa} for numerical bootstrap with defects or boundaries). 
Studying CFTs on space-time with non-trivial geometry should give us more information about the properties of the CFT.
 For example, there have been numerous studies in two dimensions, where the modular invariance puts strong constraints on the spectrum. %The quantum field theories on non-orientable space may also contain important information on the symmetry protected phase of matter.

In this paper, we use the conformal bootstrap program to solve CFTs on $d$-dimensional real projective spaces. A real projective space is defined by an involution
\begin{align}
\vec{x} \to - \frac{\vec{x}} {\vec{x}^2} \label{involution}
\end{align}
on the flat Euclidean space $R_d$ parametrized by a vector $\vec{x}$, and it preserves the $SO(1,d)$ subgroup of the original (Euclidean) conformal symmetry $SO(1,d+1)$. Alternatively, one may conformally map it to the Lorentzian cylinder $R_t \times S_{d-1}$, where the involution is given by the CPT transformation 
$(t,\vec{\Omega}) \to (-t,-\vec{\Omega})$, 
where $\vec{\Omega}$ is a unit vector on $R_d$ parametrizing $S_{d-1}$. When $d$ is even, the real projected space is not orientable. See fig 1 for illustration.

\begin{center}
\begin{figure}[h!!]
  \centering
  \includegraphics[width=8cm]{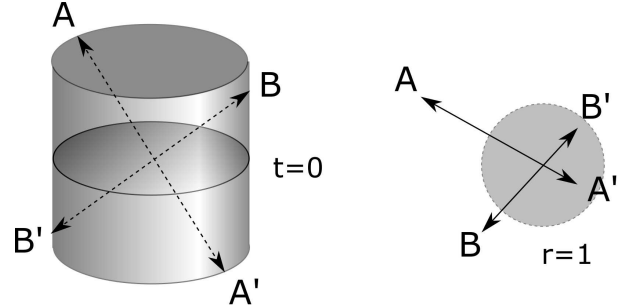}
  \caption{Involutions used in two-dimensional real projective spaces: The left panel describes the involution $(t,\vec{\Omega}) \to (-t,-\vec{\Omega})$ on the Lorentzian cylinder while the right panel describes the involution $\vec{x} \to - \frac{\vec{x}} {\vec{x}^2}$ on the Euclidean space. The fundamental domain can be chosen $t\ge0$ or $r^2 = \vec{x}^2 \ge 1$.}
  \label{fig:1}
\end{figure}
\end{center}

More recently, we found that CFTs on real projective spaces have a surprising application to constructing bulk local operators in the context of holography \cite{Verlinde:2015qfa}\cite{Miyaji:2015fia}\cite{Nakayama:2015mva}. There the building block of CFTs on real projective spaces, so-called Ishibashi states \cite{Ishibashi:1988kg} can be mapped to bulk local operators in the dual AdS space-time. This observation gives a geometrical interpretation of the crosscap conformal blocks in terms of correlation functions in AdS space-time \cite{Nakayama:2015mva}.

In two dimensions,  CFTs on a projective space have important applications to unoriented string theory. They give a worldsheet realization of orientifold planes. In rational CFTs, one may use the conformal bootstrap equation on the projective space or modular bootstrap equation on the Klein bottle and M\"obius strip to solve the crosscap states. This is possible with the help of the enhanced Virasoro symmetry with null vectors \cite{Fioravanti:1993hf}. 

So far, there has been no rigorous way to use an analogue of modular bootstrap  in higher dimensional CFTs. In this paper, we, then, resort to conformal bootstrap equations on real projective spaces to derive one-point functions. Our input parameters are conformal data on a flat Euclidean space. We study the critical Ising model and the Lee-Yang model as our examples, but our method has wider applicability to the other non-trivial CFTs as long as the conformal data on the flat Euclidean space are sufficiently well-known.

Three dimensional CFTs are notoriously hard to solve analytically, but the recent development in numerical conformal bootstrap allows us to obtain conformal data with sufficiently high precision. We use these data to numerically solve the conformal bootstrap equations on real projective spaces. We check the rapid convergence of our bootstrap program in two-dimensions from the exact solutions available, and we estimate that our systematic error on the numerically solved one-point functions of the critical Ising model on a three-dimensional real projective space is less than one percent.  As far as we are aware, our result is the first prediction of the one-point functions of the critical Ising model on a three dimensional real projective space.

%The organization of the rest of the paper is as follows. In section \ref{sec:2}, we study the conformal bootstrap equations on a real projective space and propose a numerical method to solve them. In section \ref{sec:3}, we apply  our program
% to the two-dimensional critical Ising model and Lee-Yang model to check the rapid convergence and numerical precision. In section \ref{sec:4}, we use our program to predict solutions of the critical Ising model and Lee-Yang model on three dimensional projective space. In section \ref{sec:5}, we present future directions to be pursued.
%\section{Crosscap conformal bootstrap equation and numerical approach}\label{sec:2}
%A real projective space is defined by the identification of an involution \eqref{involution} on the flat Euclidean space. 
Let us consider a CFT on a $d$-dimensional real projective space. One-point functions of scalar primary operatories $O_{i}$ with conformal dimension $\Delta_i$ are given by
\begin{align}
\langle O_{i}(\vec{x}) \rangle  = \frac{A_{i}}{(1+\vec{x}^{2})^{\Delta_i}} .
\end{align}
From the rotational invariance, higher spin operators do not possess non-zero one-point functions. Once $A_i$ are known in addition to conformal data on the flat Euclidean space, the CFT on real projective spaces is completely solved  in principle.

Two-point functions can be parameterized as
\begin{align}
& \langle O_{1}(\vec{x}_1) O_{2}(\vec{x}_2) \rangle \cr
=& \frac{(1+\vec{x}^2_1)^{\frac{-\Delta_1 + \Delta_2}{2}}   (1+\vec{x}^2_2)^{\frac{-\Delta_2+\Delta_1}{2}}}{ (\vec{x}_1 - \vec{x}_2)^{2 (\frac{\Delta_1+\Delta_2}{2})}} G_{12}(\eta) \ ,
\end{align}
where 
$\eta =  \frac{(\vec{x}_1 - \vec{x}_2)^2}{(1+\vec{x}_1^2)(1+\vec{x}_2^2)}$ is the crosscap crossratio. 
By using the conformal invariance and OPE $O_1 \times O_2 = \sum_i C_{12i} O_i$ in the $\vec{x}_1 \to \vec{x}_2$ limit (i.e. $\eta \to 0$ or $A \to B$ in fig 1), one can derive the conformal partial wave decomposition \cite{NO}:
\begin{align}
G_{12}(\eta) = \sum_i C_{12i} A_i \eta^{\Delta_i/2} \cr 
 {}_2F_1 \left(\frac{\Delta_1 - \Delta_2 + \Delta_i}{2},\frac{\Delta_2-\Delta_1+\Delta_i}{2}; \Delta_i + 1-\frac{d}{2};\eta\right) \ . \label{cpwd}
\end{align}
We are going to fix the overall normalization such that for the identity operator $A_1 = 1$, and $C_{ij1} = \delta_{ij}$. Note that the sum is taken only over scalar primary operators. 
The similar expression for disk two-point functions was derived in \cite{McAvity:1995zd}. 
In relation to the holography, each term in the sum corresponds to the bulk-boundary three-point functions in AdS space-time \cite{Kabat:2012av}\cite{Kabat:2015swa}\cite{NO}.

Now we may also expand the two-point functions by using the mirror channel OPE $\vec{x}_1 \to -\frac{\vec{x}_2}{\vec{x}_2^2}$ (i.e. $\eta \to 1$ or $A\to B'$ in fig 1). Assuming that $O_i$ transforms trivially under the involution (i.e. $O_i(\vec{x}) \sim \Gamma_{ij}O_j(-\frac{\vec{x}}{\vec{x}^2})$ with unit $\Gamma_{ij}$), we have the crossing equation
\begin{align}
\left(\frac{1-\eta}{\eta^2}\right)^{\frac{\Delta_1+\Delta_2}{6}} G_{12}(\eta) = \left(\frac{\eta}{(1-\eta)^2}\right)^{\frac{\Delta_1+\Delta_2}{6}} G_{12}(1-\eta) \ .
\label{cross} \end{align}
This will be called the crosscap bootstrap equation.
 A generalization to the case in which $O_i$ transforms non-trivially under the involution is straightforward, but we will not use it in this paper.
Our program is to solve \eqref{cross} numerically to determine $A_i$ with a given conformal data $\Delta_{i}$ and $C_{12i}$ on the flat Euclidean space. 

One way to numerically solve the crosscap bootstrap equation is to truncate the sum over $i$ and demand that \eqref{cross} hold around the symmetric point $\eta = 1/2$. While there are many options how to determine $A_i$ approximately at this point, we use the similar derivative expansions studied in \cite{Gliozzi:2013ysa}\cite{Gliozzi:2014jsa}\cite{Gliozzi:2015qsa}: we require up-to $(2M+1)$th order derivatives at $\eta = 1/2$ exactly vanish (while even number of derivatives automatically vanish). Although the method has been sometimes called ``a severe truncation" in the literature \cite{Hogervorst:2015akt}, we will see that the convergence is remarkably rapid and just a few order truncation shows a reasonably accurate estimate of the real projective space one-point functions in exactly solved examples as we will see  below. We note that in some cases like a free boson CFT, the sum is actually over a finite set in the crosscap bootstrap equation and the truncation is then exact. This is  in sharp contrast with the conformal bootstrap equations on a flat Euclidean space in which the infinite sum is always necessary.

On a theoretical basis of this truncation approach, we note the following.
If the coefficients $C_{12i} A_i$ in \eqref{cpwd} were all positive, which is indeed the case for the exactly solvable two-dimensional critical Ising model (see supplemental material), one could use the Hardy-Littlewood type asymptotic analysis to show that the convergence of the conformal partial wave decomposition \eqref{cpwd} is exponentially fast in line with the studies of \cite{Pappadopulo:2012jk}\cite{Kim:2015oca}\cite{Rychkov:2015lca} for four-point functions on a Euclidean space. In particular, only including the operators  $\Delta_i< \Delta_*$ in the summation  would cause the exponentially suppressed error of order $2^{-\frac{\Delta_*}{2}}$  in $G_{12}(\eta=1/2)$ as $\Delta_* \to \infty$ . While we unfortunately do not know the positivity of $C_{12i} A_i$ for generic CFTs, the rapid convergence we will nevertheless see may imply that no dangerous cancellation, which could ruin the asymptotic analysis, is happening.

%\section{Two-dimensional tests}\label{sec:3}
In order to test the efficiency of our numerical conformal bootstrap program, we begin our studies in two-dimensions. Rational CFTs on a two-dimensional real projective space can be exactly solved, and the Lee-Yang model and the critical Ising model that we examine are such examples. The former is an example of non-unitary CFTs, but in our bootstrap program, unitarity does not play a crucial role.

We first study a two-point function $\langle \sigma(\vec{x}_1) \sigma(\vec{x}_2) \rangle $ of the spin operator $\sigma$ in the Lee-Yang model on a real projective space. In terms of the global conformal block, the scalar OPE is given by 
\begin{align}
\sigma \times \sigma = \sigma + 1 + \epsilon' + \epsilon'' + \epsilon^3 + \cdots
\end{align}
with the conformal dimensions $\Delta_{\sigma} = -0.4$ and $\Delta_i = -0.4, 0,  4, 7.6, 11.6, 12 \cdots $.

In the crosscap bootstrap equation, we truncate the sum over the first $(n+1)$th scalar exchange (e.g. for $n=2$, we include $\sigma$,$1$, and $\epsilon'$) and demand that up to $(2n+1)$-th derivatives of the crosscap bootstrap equation at $\eta=1/2$ vanish. The resulting approximate solution of the crosscap conformal bootstrap equation for $C_{\sigma \sigma \sigma} A_\sigma$ and $C_{\sigma \sigma \epsilon'} A_{\epsilon'}$   are summarized in table \ref{table:1}. We see that already at $n=2$, the error in $C_{\sigma \sigma \sigma}A_{\sigma}$ is one percent, and the convergence to the exact solution (see supplemental material) is quite fast.
\begin{table}[htbp]
\resizebox{8.8cm}{!}{
\begin{tabular}{|c||c|c|c|c|c|c|}
\hline
 &Exact & (2,0) & (3,0) & (4,0) & (5,0)  \\ \hline \hline
$C_{\sigma \sigma \sigma} A_\sigma$ &$-3.9338$&$-3.9101$&$-3.9364$&$-3.9345$&$-3.9342$\\ \hline
$C_{\sigma \sigma \epsilon'} A_{\epsilon'}$ &$-0.01818$&$-0.02012$&$-0.01795$&$-0.01811$&$-0.01814$ \\ \hline
\end{tabular}}
\caption{Solutions of crosscap  bootstrap equation with various truncations in two-dimensional Lee-Yang model. With the notation $(n,0)$, we truncate the sum up to $n+1$ scalar operators including the identity operator.}
\label{table:1}
\end{table}

Similarly, we can study a two-point function $\langle \sigma(\vec{x}_1) \sigma(\vec{x}_2) \rangle $ of the spin operator in the two-dimensional critical Ising model on a real projective space. In terms of the global conformal block, the scalar OPE is given by
\begin{align}
\sigma \times \sigma = 1 + \epsilon + \epsilon' + \epsilon'' + \epsilon^3 + \cdots
\end{align}
with the conformal dimensions $\Delta_{\sigma} = 0.125$ and $\Delta_i =  0, 1,  4, 8,9,12, \cdots$.
The approximate solutions of the crosscap  bootstrap equation are summarized in table \ref{table:2}. Again, we see the convergence is remarkably rapid with increasing $n$. 
\begin{table}[htbp]
\resizebox{8.8cm}{!}{
\begin{tabular}{|c||c|c|c|c|c|c|}
\hline
 &Exact & (2,0) & (3,0) & (4,0) & (5,0)  \\ \hline \hline
$C_{\sigma \sigma \epsilon} A_\epsilon$ &$0.20711$&$0.20407$&$0.20757$&$0.20693$&$0.20710$\\ \hline
$C_{\sigma \sigma \epsilon'} A_{\epsilon'}$ &$0.01563$&$0.01702$&$0.01539$&$0.01572$&$0.01563$ \\ \hline
\end{tabular}}
\caption{Solutions of crosscap bootstrap equation with various truncations in two-dimensional critical Ising model.}
\label{table:2}
\end{table}

So far, we have used the conformal data on a flat Euclidean space as a given input. However, in harmony with a spirit of \cite{Gliozzi:2015qsa}, one may determine the conformal data self-consistently from the crosscap bootstrap equation. This is done by requiring that up to $(2(n+2m)+1)$-th derivative vanish given $n$ input conformal data. The condition determines $m$ additional conformal data. 
%Note that one cannot reduce $n$ to be zero because in principle there are many solutions  of the most general conformal bootstrap equations and one may need some input parameters.

In table \ref{table:3}, we summarize our attempt to determine conformal data of the two-dimensional critical Ising model on the flat Euclidean space self-consistently from the crosscap bootstrap equation. We typically see that the predicted conformal dimensions are larger than the known exact values, but this may be expected because they represent contributions of all the higher dimensional operators than the truncated ones. We also observe that irrespective of the precision of the predicted conformal dimensions, the obtained one-point functions become better approximated in all examples we have studied.

\begin{table}[htbp]
\resizebox{8.8cm}{!}{
\begin{tabular}{|c||c|c|c|c|c|c|}
\hline
 &Exact & (1,2) & (2,1) & (2,2) & (3,1)  \\ \hline \hline
$C_{\sigma \sigma \epsilon} A_\epsilon$ &$0.20711$&$0.21499$&$0.20653$&$0.20696$&$0.20712$\\ \hline
$C_{\sigma \sigma \epsilon'} A_{\epsilon'}$ &$0.01563$&$0.01248$&$0.01593$&$0.01571$&$0.01558$ \\ \hline
spectrum & &$\Delta_{\epsilon'} = 4.4$& $\Delta_{\epsilon''} = 9.1$&$\Delta_{\epsilon''} = 8.4$&$\Delta_{\epsilon^3} = 10.8$ \\ \hline
predictions  & &$\Delta_{\epsilon''} = 11.1$&& $\Delta_{\epsilon^3} = 12.2$& \\ \hline
\end{tabular}}
\caption{Solutions of crosscap bootstrap equation with additional self-consistent predictions of the spectrum in two-dimensional critical Ising model. With the notation $(n,m)$  we truncate the sum up to $n+1$ scalar operators including the identity operator, and predict $m$ additional conformal data.}
\label{table:3}
\end{table}

%\section{Three-dimensional predictions}\label{sec:4}
Given the success of our numerical conformal bootstrap program on a real projective space in two-dimensions, we now apply the same idea to three-dimensional CFTs. This will result in the first predictions of one-point functions of interacting CFTs on a three-dimensional real projective space.

Let us begin with the Lee-Yang model in three dimensions. The conformal data, in particular the dimensions of irrelevant operators, is much less known in this model, but we use the recent result obtained from the self-consistent numerical bootstrap for four-point functions on a Euclidean space in \cite{Gliozzi:2014jsa}:
\begin{align}
\Delta_{\sigma} = 0.235 \ , \ \  \Delta_{\epsilon'} = 4.75 \ . 
\end{align}
%with the scalar OPE 
%\begin{align}
%\sigma \times \sigma = \sigma + 1 + \epsilon' + \cdots
%\end{align}

In table \ref{table:4} we show the numerical solutions of the truncated crosscap bootstrap equation. 
Since we have much less conformal data than in two-dimensions, we have also tried to improve our estimate by including the self-consistent predictions of higher dimensional operators as shown in table \ref{table:4}. 
As in two-dimensions, we do not expect that the self-consistent prediction gives the precise conformal dimensions, but we expect the one-point functions become better approximated.

\begin{table}[htbp]
\resizebox{8.8cm}{!}{
\begin{tabular}{|c||c|c|c|c|c|c|}
\hline
 &(2,0) & (1,1) & (1,2) & (2,1) & (2,2)  \\ \hline \hline
$C_{\sigma \sigma \sigma} A_\sigma$ &$-1.369$&$-1.352$&$-1.3587$&$-1.3656$&$-1.3650$\\ \hline
$C_{\sigma \sigma \epsilon'} A_{\epsilon'}$ &$-0.00613$&$-0.00409$&$-0.00485$&$-0.00561$&$-0.00551$\\ \hline
spectrum & &$\Delta_{\epsilon'} = 5.5$& $\Delta_{\epsilon'} = 5.1$&$\Delta_{\epsilon''} = 9.6$&$\Delta_{\epsilon''} = 8.9$ \\ \hline
predictions & &&$\Delta_{\epsilon''} = 11.3$ & & $\Delta_{\epsilon^3}=12.4$ \\ \hline
\end{tabular}}
\caption{Solutions of crosscap bootstrap equation with additional self-consistent predictions of the spectrum in three-dimensional Lee-Yang model.}
\label{table:4}
\end{table}

Finally, we numerically solve the crosscap bootstrap equation in the critical Ising model in three dimensions. As for the conformal data on the flat Euclidean space in particular for higher dimensional scalar operators, we use two sets of parameters computed by the self-consistent truncated bootstrap solutions of boundary two-point functions  \cite{Gliozzi:2015qsa} and the extremal functional method  \cite{ElShowk:2012hu} in numerical bootstrap of the four-point functions on the flat Euclidean space.

The one from the truncated boundary bootstrap is
\begin{align}
\text{Set A:}& \ \Delta_{\sigma} = 0.518154, \ \Delta_{\epsilon} = 1.41267, \  \cr& \ \Delta_{\epsilon'} = 3.8303, \
\Delta_{\epsilon''} = 7.27, \ \Delta_{\epsilon^3} = 12.90.
\end{align}

The one from the extremal functional method implemented by Juliboots \cite{Paulos:2014vya} with $n_{\text{max}}=21$, $m_{\text{max}}=1$ is
\begin{align}
\text{Set B:}& \ \Delta_{\sigma} = 0.518151, \ \Delta_{\epsilon} = 1.412658, \ \cr& \ \Delta_{\epsilon'} = 3.83032, 
\Delta_{\epsilon''} = 6.9905, \ \Delta_{\epsilon^3} = 10.83,
\cr  &\ \Delta_{\epsilon^4} = 15.22, \ \Delta_{\epsilon^5} = 21.07 \ . \label{setB}
\end{align}
Here, the extremal functional includes order $\frac{n_{\text{max}}^2}{2}$ operators, but the dimensions of scalar operators $\Delta > 10$ are sensitive to $n_{\text{max}}$ used. This will be the one of the main sources of the systematic error in our program. We have also checked the sensitivity with the implementation by SDPB \cite{Simmons-Duffin:2015qma}, but the results do not improve much.

Based on these conformal data, we compute the approximate solutions of the crosscap bootstrap equation as shown in table \ref{table:5}. 
For comparison, we have also included our self-consistent predictions of the operator spectrum with $n=2$ and $m=2$. The result is $\Delta_{\epsilon''} = 7.60$ and  $\Delta_{\epsilon^3} = 10.7$,  
which may be compared with the above self-consistent truncated solutions from the boundary bootstrap.
\begin{table}[htbp]
\resizebox{8.8cm}{!}{
\begin{tabular}{|c||c|c|c|c|c|c|}
\hline
 &(2,0) & (4,0)$_{\mathrm{A}}$ & (4,0)$_{\mathrm{B}}$ & (6,0)$_{\mathrm{B}}$ & (2,2)  \\ \hline \hline
$C_{\sigma \sigma \epsilon} A_\epsilon$ &0.690&0.7015&0.7022&0.70197&0.7006\\ \hline
$C_{\sigma \sigma  \epsilon'} A_{\epsilon'}$ &0.054&0.0475&0.0470&0.04714&0.0480 \\ \hline
\end{tabular}}
\caption{Solutions of crosscap conformal bootstrap equation for the critical Ising mode in three dimensions.}
\label{table:5}
\end{table}

In order to check the stability of the truncation and estimate the systematic error with respect to the less reliable spectrum of higher dimensional operators, we have repeated our analysis by artificially changing the values of $\Delta_{\epsilon^4}$ and  $\Delta_{\epsilon^5}$ in set B, ranging between $(14,16)$ and $(18, 25)$ respectively. The predictions of $C_{\sigma \sigma \epsilon} A_\epsilon$ and $C_{\sigma \sigma  \epsilon'} A_{\epsilon'}$ then fluctuate around $0.70197 \pm 0.00008 $ and $0.04714\pm 0.00007$. 
As in \cite{Gliozzi:2015qsa}, we may regard the fluctuation as a rough estimate of the systematic error of our approach, which is small and comparable to the truncation error we saw in the two-dimensional critical Ising model.

Combining the known OPE coefficients (see e.g. \cite{El-Showk:2014dwa}: the number we use below is obtained from the output of Juliboots with $n_{\text{max}}=21$, $m_{\text{max}}=1$)
\begin{align}
C_{\sigma\sigma\epsilon} = 1.0518 \ , \ \ C_{\sigma\sigma \epsilon'} = 0.0530 \ , 
\end{align}
we finally predict
\begin{align}
A_{\epsilon} = 0.667(2)  \ , \ \  A_{\epsilon'} = 0.896(5) 
\end{align}
for the three-dimensional critical Ising model on a  real projective space.
From the numerical convergence and the comparison with the experience in two-dimensional cases, we estimate that our systematic error is  less than one percent.

%\section{Discussions}\label{sec:5}
In this paper, we have proved that the conformal bootstrap is useful not only on  flat Euclidean spaces, but also on real projective spaces. We would like to conclude the paper with future directions to be pursued.

First of all, we have studied the simplest models i.e. the Lee-Yang model and the critical Ising model, but our method can be easily generalized to other models such as critical $O(n)$ models. Assuming the simplest involution that commutes with $O(n)$ symmetry, the crosscap bootstrap equations for $\langle \sigma_i(\vec{x}_1)\sigma_j(\vec{x}_2) \rangle$ take the same form since only the $O(n)$ singlet scalar operator contributes to the sum, so one may solve the equations in the same way as we have done in this paper. The only difference is the input conformal data, which can be obtained along the line of \cite{Kos:2013tga}\cite{Kos:2015mba} from the conformal bootstrap.

Another interesting future direction is to study other correlation functions.  Indeed, since we have determined the projective space one-point function only from  $\langle \sigma(\vec{x}_1)\sigma(\vec{x}_2) \rangle$, the study of the other correlation functions yields strong consistency checks. More ambitiously we may imagine that we would be able to predict conformal data on the flat Euclidean space such as $C_{\sigma\sigma\epsilon}/C_{\epsilon\epsilon\epsilon}$ from the consistency conditions of multi-correlator crosscap bootstrap equations.

Finally, it would be interesting to check our predictions from the direct studies of the correlation functions on real projective spaces e.g. from epsilon expansions, Hamiltonian truncations, and Monte-Carlo simulations. On computers, there is no obstruction to study statistical models on  real projective spaces while experimental tests may require some ingenuity.

%\section*{Acknowledgements}
The author would like to thank  T.~Ohtsuki, H.~Ooguri and S.~Rychkov for discussions and suggestions. This work is supported by the World Premier International Research Center Initiative (WPI Initiative), MEXT. 

%T.O. is supported by JSPS Research Fellowships for Young Scientists and the Program for Leading Graduate Schools, MEXT.

\appendix

\section{Exact bootstrap solutions in two-dimensions}
In this appendix, we will present the exact solutions of the crosscap bootstrap equations with the Virasoro symmetry in two-dimensional Lee-Yang model and the critical Ising model.
In rational CFTs, we only need a finite number of Virasoro conformal blocks to solve the crosscap bootstrap equations. Furthermore due to null vectors in the Virasoro algebra, one may derive explicit forms of the Virasoro crosscap conformal block in the minimal models. Then the crosscap bootstrap equations are reduced to algebraic equations that can be easily solved.

In the Lee-Yang model, we have the Virasoro OPE
\begin{align}
[\sigma] \times [\sigma] = [1] + [\sigma] \ .
\end{align}
Accordingly,  the two-point function of the spin operator $\sigma$ on the real projective space can be decomposed into two Virasoro blocks:
\begin{align}
G_{\sigma\sigma}(\eta) &= (1-\eta)^{1/5} {}_2F_1(\frac{2}{5},\frac{3}{5};\frac{6}{5};\eta) \cr 
&+ c_{\mathrm{LY}} \eta^{-1/5} (1-\eta)^{1/5} {}_2F_1(\frac{1}{5},\frac{2}{5};\frac{4}{5};\eta) \ ,
\end{align}
where 
\begin{align}
c_{\mathrm{LY}} &= \frac{\Gamma(\frac{1}{5})(\Gamma(\frac{2}{5})\Gamma(\frac{3}{5})-\Gamma(-\frac{1}{5})\Gamma(\frac{6}{5}))}{\Gamma(\frac{3}{5})\Gamma(-\frac{1}{5}) \Gamma(\frac{4}{5})} \cr
&= - \frac{2^{1/5} \sqrt{5(5+2\sqrt{5})\pi}\Gamma(\frac{1}{10})}{\Gamma(-\frac{1}{5})^2 } \cr
&= -3.9338 
\end{align}
is determined from the crossing symmetry.
In terms of the global conformal block, we have the decomposition
\begin{align}
G_{\sigma\sigma}(\eta) & = 1 - \frac{\eta^2}{55} {}_2F_1(2,2;4;\eta)  + \frac{\eta^6}{596750} {}_2F_1(6,6;12;\eta) + \cdots \cr
& +c_{\mathrm{LY}} ( \eta^{-\frac{1}{5}}  {}_2F_1(-\frac{1}{5},-\frac{1}{5};-\frac{2}{5};\eta)  \cr
&+ \frac{\eta^{\frac{19}{5}}}{7410} {}_2F_1(\frac{19}{5},\frac{19}{5};\frac{38}{5};\eta) + \cdots )
\end{align}
The coefficients in front of each block were quoted as the exact results in table \ref{table:1}.

In the critical Ising model, we have the Virasoro OPE
\begin{align}
[\sigma] \times [\sigma] = [1] + [\epsilon]
\end{align}
and the two-point function  of the spin operator $\sigma$ on the real projective space can be decomposed into
\begin{align}
G_{\sigma\sigma}(\eta) &= (1-\eta)^{3/8} {}_2F_1(\frac{3}{4},\frac{1}{4};\frac{1}{2};\eta) \cr
 &+c_{\mathrm{I}} \eta^{1/2} (1-\eta)^{3/8} {}_2F_1(\frac{3}{4},\frac{5}{4};\frac{3}{2};\eta) ,
\end{align}
where $c_{\mathrm{I}}= \frac{\sqrt{2}-1}{2} = 0.20711$. This number was also obtained from the modular bootstrap in \cite{Fioravanti:1993hf}.

Note that the two-point function can be expressed in terms of more elementary functions by using
\begin{align}
 {}_2F_1(\frac{3}{4},\frac{5}{4};\frac{3}{2};\eta) = \frac{\sqrt{2} \sqrt{1-\sqrt{1-\eta}}}{\sqrt{1-\eta}\sqrt{\eta}}  \cr
{}_2F_1(\frac{3}{4},\frac{1}{4};\frac{1}{2};\eta)  = \frac{\sqrt{1+\sqrt{1-\eta}}}{\sqrt{2}\sqrt{1-\eta}} \ .
\end{align}

In terms of the global conformal block, we have the decomposition
\begin{align}
G_{\sigma\sigma}(\eta) & = 1 + \frac{\eta^2}{64} {}_2F_1(2,2;4;\eta)  + \frac{9\eta^4}{40960} {}_2F_1(4,4;8;\eta) + \cdots \cr
& +c_{\mathrm{I}} ( \eta^{\frac{1}{2}}  {}_2F_1(\frac{1}{2},\frac{1}{2};1;\eta) + \frac{\eta^{\frac{9}{2}}}{16384} {}_2F_1(\frac{9}{2},\frac{9}{2};9;\eta) + \cdots )
\end{align}
The coefficients in front of each block were quoted as the exact results in table \ref{table:2}.

In the above solutions of the crosscap bootstrap equations, 
Euler's transformation 
\begin{align}
&{}_2F_1(a,b;c;\eta) = (1-\eta)^{c-a-b} {}_2F_1(c-a,c-b;c;\eta) \cr
 & = \frac{\Gamma(c)\Gamma(c-a-b)}{\Gamma(c-a)\Gamma(c-b)} {}_2F_1(a,b;a+b+1-c;1-\eta) \cr
&+ \frac{\Gamma(c)\Gamma(a+b-c)}{\Gamma(a)\Gamma(b)} (1-\eta)^{c-a-b}  \cr
&\times {}_2F_1 (c-a,c-b;1+c-a-b;1-\eta) \ 
\end{align}
may be useful.

\end{document}